\documentclass[12pt]{article}
\usepackage[utf8]{inputenc}
\usepackage[english]{babel}
\usepackage{amsmath}
\numberwithin{equation}{section}
\usepackage{amsthm,amssymb}
\usepackage{bm}
\usepackage{amsfonts}
\usepackage[margin=1in,includefoot]{geometry}
\usepackage{fancyvrb}
\usepackage{dsfont}
\usepackage{mathdots}
\usepackage{mathtools}
\usepackage{mathrsfs}
\usepackage{float}
\usepackage{caption}
\usepackage{pgfgantt}
\usepackage{indentfirst}
\usepackage{tikz}
\usepackage[font=small,labelfont=bf]{caption}
\usepackage[nottoc]{tocbibind}
\usepackage[toc]{appendix}
\usepackage{algorithm}
\usepackage{multirow}
\usepackage{titlesec}
\usepackage[affil-it]{authblk}

\allowdisplaybreaks

\newcommand{\1}{\mathbf{1}}

\author[1]{Yiping Guo}
\author[2]{Johnny Siu-Hang Li}
\affil[1]{\normalsize Department of Statistics and Actuarial Science, University of Waterloo}
\affil[2]{\normalsize Department of Finance, The Chinese University of Hong Kong}

\usepackage[round]{natbib}
\title{Fast Estimation of the Renshaw-Haberman Model and Its Variants}
\date{}

\allowdisplaybreaks

\begin{document}
\maketitle
\begin{abstract}
\noindent 
In mortality modelling, cohort effects are often taken into consideration as they add insights about variations in mortality across different generations. Statistically speaking, models such as the Renshaw-Haberman model may provide a better fit to historical data compared to their counterparts that incorporate no cohort effects. However, when such models are estimated using an iterative maximum likelihood method in which parameters are updated one at a time, convergence is typically slow and may not even be reached within a reasonably established maximum number of iterations. Among others, the slow convergence problem hinders the study of parameter uncertainty through bootstrapping methods. In this paper, we propose an intuitive estimation method that minimizes the sum of squared errors between actual and fitted log central death rates. The complications arising from the incorporation of cohort effects are overcome by formulating part of the optimization as a principal component analysis with missing values. Using mortality data from various populations, we demonstrate that our proposed method produces satisfactory estimation results and is significantly more efficient compared to the traditional likelihood-based approach.

\end{abstract}
\textbf{Keywords}: 
Cohort effects; Missing values; Principal component analysis


\section{Introduction} 

One important concept in modeling and management of longevity risk is cohort effects. In the context of longevity risk, cohort effects refer to the impact of a person's birth year or generation on their health and mortality outcomes. The significance of cohort effects has long been recognized by demographers \citep{Cohort-1982-Hobcraft, Cohort-1990-Wilmoth} and actuaries \citep{UK-2004-Willets}. 

Cohort effects can be attributed to various factors such as changes in lifestyle, medical advancements, etc. Their strength varies across geographical regions, although it is widely acknowledged that they are particularly strong in the United Kingdom, where the ``golden generation'' who were born in the early 1930s experienced significantly higher mortality improvement. It is important to note that cohort effects are not purely historical. For example, starting in July 2023, New Zealand bans the sale of tobacco products to individuals born on or after January 1, 2009. The ban is expected to result in cohort effects in mortality improvement for New Zealanders, as their younger generations are not exposed to the negative effects of tobaccos. 

To incorporate cohort effects into stochastic mortality modeling, \cite{RH-2006-Renshaw} extend the seminal work of \cite{LC-1992} to develop the Renshaw-Haberman model. It adds to the original Lee-Carter model a bi-linear term, which captures the variation of mortality across years-of-birth and the interaction between such variation with age. It is also closely connected to the classical age-period-cohort (APC) model \citep{Cohort-1982-Hobcraft}, as it degenerates to the APC model when some of its age-specific parameters are eliminated. When estimated to historical mortality data, the model is able to absorb part of the remaining variation that is not captured by models with age and period (time-related) effects only, leaving residuals that exhibit a more random pattern. Recently, the Renshaw-Haberman model has been generalized to incorporate socioeconomic differences in mortality \citep{Mdiff-2014-Villegas}, making it applicable to an even wider range of insurance and pension applications. 

In the literature, including the original work of \cite{RH-2006-Renshaw}, the Renshaw-Haberman model is often estimated with maximum likelihood (ML). When fitting the Renshaw-Haberman with ML, a log-likelihood function is derived on the basis of a distributional assumption, typically Poisson, made on observed death counts; then, parameter estimates are obtained by maximizing the log-likelihood function. Given that the Renshaw-Haberman model has a large number of parameters, the maximization is customarily performed with an iterative Newton-Raphson method, in which parameters are updated one batch at a time. Unfortunately, ML estimation for the Renshaw-Haberman model is slow and sometimes unstable. Depending on the dataset in question, the iterative algorithm may not even converge given the desired convergence criterion. This problem is noted by a number of researchers, including \cite{Comparison-2009-Cairns, Comparison-2011-Cairns} and \cite{RH-2009-Haberman, Comparison-2011-Haberman}. 

While a slow convergence might be acceptable is model estimation is a one-off task, it may render applications that require repeated model estimation time-prohibitive. Such applications include the following.
\begin{itemize}
    \item {\it Assessment of parameter uncertainty via bootstrapping}
    
    Any model-based mortality projection is subject to parameter uncertainty, as the parameters used for extrapolating future death rates are estimates rather than exact. One way to gauge parameter uncertainty is bootstrapping \citep{brouhns2005bootstrapping, d2012stratified, BootLC-2006-Koissi}. In a bootstrap, a large number pseudo datasets are generated by, for example, resampling residuals (residual bootstrapping); then, the model is re-estimated using the pseudo data sets. The procedure results in empirical distributions of model parameters, from which parameter uncertainty can be inferred. The bootstrapping procedure involves a large number of (re-)estimations, and cannot be executed in practice if the estimation is slow. 

    \item {\it Calculation of Solvency Capital Requirements}

    Under Solvency II, solvency capital requirement (SCR) is based on the Value-at-Risk at a 99.5\% confidence level over a one-year horizon \citep{zhou2014modeling}. In lieu of the prescribed standard formula, an insurer may opt to calculate SCR by simulating from an approved internal model. Taking re-calibration risk\footnote{Re-calibration risk arises because model parameter estimates may become different if the model in question is fitted to an updated data set.} \citep{cairns2013robust} into account, the simulation procedure for estimating longevity Value-at-Risk encompasses the following steps: (1) simulate $M_1$ mortality scenarios in one year from a model that is fitted to historical data; (2) for each mortality scenario, re-estimate the model to an updated dataset that includes the simulated mortality scenario, and use the re-estimated model to simulate $M_2$ sample paths of mortality (for year 2 and beyond), from which the expected value of the liability at the end of year 1 can be calculated. Step (2) yields a distribution of liabilities at the end of year 1, which can be used to infer the 99.5\% Value-at-Risk. Typically, $M_1$ is large, so that the procedure includes a large number of model re-estimations. 

    \item {\it Identification of ultimate mortality improvement rates}

    In recent years, two-dimensional mortality improvement scales have been promulgated by major actuarial professional organizations \citep[see, e.g.,][]{soa2021MI}. A two-dimensional mortality improvement scale is composed of relatively high short-term scale factors, which are blended into lower long-term (ultimate) scale factors through an interpolative mid-term scale. One possible way to estimate the ultimate scale factors is to fit a parametric model to absorb all transient period and cohort effects that are present in the historical data, leaving a long-term pattern from which the ultimate scale factors can be inferred \citep{li2020drivers}. This method requires the modeller to experiment different model structures, some of which, ideally, include multiple age-cohort interaction terms. A slow convergence rate plagues the use of this method; in particular, it hinders the consideration of models with additional age-cohort interaction terms.
    
\end{itemize}


So far as we aware, two attempts have been made to mitigate the estimation issues of the Renshaw-Haberman model. The first attempt is made by \cite{RH-2006-Renshaw}, who consider a number of restricted versions of the Renshaw-Haberman model which may take less time to estimate given that they have fewer free parameters. Most notably, they propose the H1 model, which still incorporates cohort effects but assumes that such effects do not interact with age. The second attempt is made by  \cite{RH-2015-Hunt}, who argue that the problem of slow convergence is due possibly to an approximate identification issue that is applicable to the Renshaw-Haberman model. To mitigate the issue, they recommend imposing an additional parameter constraint to stabilize the estimation process and enhance algorithmic robustness. It is noteworthy that both approaches are based on a reduction in parameter space. That said, they improve estimation efficiency at the expense of goodness-of-fit to the historical data.

In this paper, we attack the problem of estimation efficiency for the Renshaw-Haberman model from a different angle. Instead of building on the commonly used maximum likelihood approach, we consider a least squares method in which parameters are estimated by minimizing the sum of squared errors between the actual and fitted log central death rates. The idea of using a least squares approach to estimate stochastic mortality models is not new. As a matter of fact, when the original Lee-Carter and Cairns-Blake-Dowd models were first proposed, the authors estimated them with least squares methods \citep{LC-1992, CBD-2006-Cairns}. 

It is not straightforward to efficiently estimate the Renshaw-Haberman model with a least squares method. This is because the model involves an additional (year-of-birth) dimension that is not orthogonal to the age and time dimensions, rendering the efficient singular value decomposition (SVD) technique that is used for fitting the original Lee-Carter model inapplicable. To overcome the optimization challenge,  we develop an alternating minimization scheme which sequentially updates one group of parameters at a time. We also formulate the update of the age-cohort component in the model as a principal component analysis (PCA) problem with missing values, so that it can be accomplished effectively using an iterative SVD algorithm. Using data from various national populations, we demonstrate that our proposed least squares method significantly outperforms the ML approach in terms of estimation efficiency, without sacrificing goodness-of-fit to historical data.

Our proposed least squares method offers several advantages over the ML approach. First, given the same convergence criterion, our proposed method takes less computation time. We argue that the improvement in estimation efficiency is due to a sharper objective function, and empirically verify this argument with a numerical experiment. Second, unlike the ML approach, our proposed method requires no distributional assumption, thereby avoiding the potential problems associated with a distribution mis-specification. Finally, our proposed method can be implemented seamlessly with the two methods that are previously proposed by \cite{RH-2006-Renshaw} and \cite{RH-2015-Hunt} to further improve estimation efficiency. 

The remainder of this paper is organized as follows. Section 2 presents an overview of the Lee-Carter model, with a focus on the estimation methods for the model that are relevant to this study. Section 3 reviews the Renshaw-Haberman model and its estimation challenges. Section 4 details our proposed method, including its motivation, theoretical support, and execution. Section 5 explains how our proposed method can be implemented simultaneously with the two methods that are previously proposed by \cite{RH-2006-Renshaw} and \cite{RH-2015-Hunt}. Section 6 documents the numerical experiments that validate the advantages of our proposed method. Finally, concluding remarks are provided in Section 7.


\section{The Lee-Carter Model}

\subsection{Specification}

This section presents a concise review of the Lee-Carter model \citep{LC-1992}, with a focus on two commonly used methods for estimating the model. We let $m_{x,t}$ be the central rate of death for age $x$ and year $t$, and $y_{x,t}:=\log(m_{x,t})$ for notational convenience. The Lee-Carter model assumes that
\begin{equation}\label{Lee-Carter}
y_{x,t}:=\log(m_{x,t})=a_x+b_xk_t+\varepsilon_{x,t},
\end{equation}
where $a_x$ and $b_x$ are age-specific parameters, $k_t$ is a time-varying index, and $\varepsilon_{x,t}$ is the error term. In the model, $a_x$ captures `age effects' (the age pattern of mortality), $k_t$ captures `period effects' (changes in the overall mortality level over time), and $b_x$ measures the interaction between age and period effects. Throughout this paper, we assume that the data set in question covers $p$ ages, $x\in [x_1,\cdots,x_p]$, and $n$ calendar years, $t\in [t_1,\cdots,t_n]$. 

The Lee-Carter model is subject to an identifiability problem. It can be shown that two parameter constraints are required to stipulate parameter uniqueness. In the literature (including the original work of \cite{LC-1992}), the following two parameter constraints are typically imposed:
\begin{equation}\label{LC-constraints}
    \sum_{x=x_1}^{x_p}b_x=1 \quad \text{and} \quad \sum_{t=t_1}^{t_n}k_t=0.
\end{equation}

\subsection{Least Squares Estimation}

In their original work, \cite{LC-1992} estimated \eqref{Lee-Carter} using a least squares approach, in which parameter estimates are chosen such that they minimize the sum of squared errors between the observed and fitted log central mortality rates. In more detail, let us rewrite the model in vector form as follows:
\begin{equation}\label{Lee-Carter Vector Form}
    \bm{y}_{t} = \bm{a}+\bm{b}k_t + \bm{\varepsilon}_t,
\end{equation}
where $\bm{y}_{t}=(y_{x_1,t},\cdots,y_{x_p,t})^T$, $\bm{a}=(a_{x_1},\cdots,a_{x_p})^T$, $\bm{b}=(b_{x_1},\cdots,b_{x_p})^T$ and $\bm{\varepsilon}_t=(\varepsilon_{x_1,t},\cdots,\varepsilon_{x_p,t})^T$. In using the least squares approach, the estimates of $\bm{a}$, $\bm{b}$ and $\bm{k}$ are obtained by solving the following optimization:
\begin{equation}\label{Lee-Carter MSE}
    \min_{\bm{a},\bm{b},\bm{k}}\sum_{x,t}(y_{x,t}-(a_x+b_xk_t))^2= \min_{\bm{a},\bm{b},\bm{k}}\sum_{t}\Vert\bm{y}_t-(\bm{a}+\bm{b}k_t)\Vert_2^2,
\end{equation}
where $\bm{k}=(k_{t_1},\cdots,k_{t_n})^T$ and $\Vert \cdot \Vert_2$ denotes the Euclidean norm (or $L^2$-norm) of a vector. When the identification constraints specified in \eqref{LC-constraints} are applied, the optimization problem specified in \eqref{Lee-Carter MSE} is equivalent to a special case of principal component analysis (PCA) with one principal component. Its solution can thus be obtained by performing a singular value decomposition (SVD) on the mean-centered log mortality data matrix, $\bm{Y}-\bar{\bm{Y}}:=(\bm{y}_{t_1}-\bar{\bm{y}},\cdots,\bm{y}_{t_n}-\bar{\bm{y}})$, where $\bar{\bm{Y}}=(\bar{\bm{y}},\cdots,\bar{\bm{y}})$ and $\bar{\bm{y}}=\frac{1}{n}\sum_{t=t_1}^{t_n} \bm{y}_t$. The solution has the following closed form:
\begin{equation}\label{Lee-Carter solution}
    \hat{\bm{a}}=\Bar{\bm{y}}, \quad \Hat{\bm{b}}=\frac{\bm{u}}{\1^T\bm{u}}, \quad \Hat{\bm{k}}= (\1^T\bm{u}) \cdot (\bm{Y}-\Bar{\bm{Y}})^T \bm{u},
\end{equation}
where $\bm{u}$ is the first left-singular vector of $\bm{Y}-\Bar{\bm{Y}}$ and $\1=(1,\cdots,1)^T$. In this solution, the term $\bm{1}^T\bm{u}$ normalizes the standard PCA solution due to the imposed constraint $\sum_xb_x=1$. Additionally, it is easy to check that the constraint $\sum_t k_t=0$ is also met. 

\subsection{Maximum Likelihood Estimation}

In contrast to the least squares approach, maximum likelihood estimation (MLE) requires a distributional assumption. Estimation of the Lee-Carter model using maximum likelihood was first accomplished by \cite{wilmoth1993computational}, who assumes that the observed death count in each age-time cell follows a Poisson distribution. We let $D_{x,t}$ be the observed number of deaths for age $x$ and year $t$, and $N_{x,t}$ be the corresponding exposure-to-risk. The method of Poisson-MLE assumes that
\begin{equation}\label{GLM}
    D_{x,t}\sim \text{Poisson}(N_{x,t}m_{x,t}), \text{ with }\log(m_{x,t})=a_x+b_xk_t.
\end{equation}
Parameter estimates are obtained by maximizing the following log-likelihood function:
\begin{equation}\label{LC-GLM-MLE}
    \ell(\bm{a},\bm{b},\bm{k})=\sum_{x,t}\left(D_{x,t}(a_x+b_xk_t)-N_{x,t}e^{a_x+b_xk_t} \right)+\text{constant}.
\end{equation}
The optimization problem can be solved via an iterative Newton-Raphson method \citep{IRWS-1979-Goodman}.


\section{The Renshaw-Haberman Model}

\subsection{Specification}

The focus of this paper is the Renshaw-Haberman model \citep{RH-2006-Renshaw}, which extends the Lee-Carter model by incorporating cohort effects. The Renshaw-Haberman model assumes that
\begin{equation}\label{RH}
    y_{x,t}:=\log(m_{x,t})=a_x+b_xk_t+c_x\gamma_{t-x}+\varepsilon_{x,t}.
\end{equation}
In the above, $\gamma_{t-x}$ is an index that is linked to year-of-birth $(t -x)$, thereby capturing cohort effects. Parameter $c_x$ captures the sensitivity of the log central death rate at each age to cohort effects. The interpretations of $a_x$, $b_x$ and $k_t$ in \eqref{RH} are the same as those in the Lee-Carter model.

The Renshaw-Haberman model is also subject to an identifiability problem. In addition to the two constraints specified in \eqref{LC-constraints}, two constraints on $c_x$ and $\gamma_{t-x}$ are needed. Following \cite{RH-2006-Renshaw}, the additional constraints we use are 
\begin{equation}\label{RH-add-constraints}
    \sum_{x=x_1}^{x_p}c_x=1, \quad \mbox{and} \quad \sum_{t-x=t_1-x_p}^{t_n-x_1}\gamma_{t-x}=0.
\end{equation}
It is worth-noting that the constraint for $\gamma_{t-x}$ may be formulated differently. For instance, as mentioned by \cite{RH-2006-Renshaw}, another possible choice is $\gamma_{t_1-x_p}=0$. We choose to use $\sum_{t-x=t_1-x_p}^{t_n-x_1}\gamma_{t-x}=0$, because it is commonly adopted in the literature \cite[e.g.,][]{Comparison-2009-Cairns} and used in the \textbf{StMoMo} package in \textbf{R} \citep{StMoMo-2015-Villegas}. The choice of the constraints makes no difference to the goodness-of-fit.

\subsection{Estimation}

Estimation of Renshaw-Haberman model is well-known to be challenging. While the Lee-Carter model can be estimated readily using a least squares approach, a parallel least squares method for estimating the Renshaw-Haberman model is not available in the literature. The least squares solution to the Renshaw-Haberman estimation problem is not easy to obtain, because the incorporation of cohort effects expands the dimension of the problem. This challenge is succinctly described by \cite{CohortBa-2018-Fung}:
\begin{quote}
    {\it ``Under the Lee–Carter original approach, one might consider modelling the crude death rate with cohort effects as follows:
$$\log(\Tilde{m}_{x,t})=\alpha_x+\beta_x\kappa_{t}+\beta_x^{\gamma}\gamma_{t-x}+\varepsilon_{x,t}.
    $$
    However the dimension of the cohort index would cause difficulty for the SVD estimation approach."}
\end{quote}

In the literature, the Renshaw-Haberman model is often estimated using maximum likelihood. Assuming Poisson death counts, the log-likelihood function for the Renshaw-Haberman model is given by
\begin{equation}\label{RH-GLM-MLE}
    \ell(\bm{a},\bm{b},\bm{k},\bm{c},\bm{\gamma})=\sum_{x,t}\left(D_{x,t}(a_x+b_xk_t+c_x\gamma_{t-x})-N_{x,t}e^{a_x+b_xk_t+c_x\gamma_{t-x}} \right)+\text{constant},
\end{equation}
where $\bm{c}=(c_{x_1},\cdots,c_{x_p})$ and $\bm{\gamma}=(\gamma_{t_1-x_p},\cdots,\gamma_{t_n-x_1})$. This objective function is maximized through an iterative Newton-Raphson method to obtain parameter estimates. Although Poisson-MLE is technically feasible for the Renshaw-Haberman model, computational efficiency represents a significant concern to users. It is widely reported that Poisson-MLE for the Renshaw-Haberman model takes a lot of iterations to converge \citep{Comparison-2009-Cairns, Comparison-2011-Cairns, RH-2009-Haberman, Comparison-2011-Haberman}. The problem is investigated more deeply by \cite{RHStart-2016-Currie}, who emphasized the importance of using appropriate starting values in the estimation process.

\subsection{Existing Methods for Expediting Estimation}

So far as we aware, there have been two major attempts to expedite estimation for the Renshaw-Haberman model. These methods are reviewed in this subsection.

\subsubsection{The H1 Model}

\cite{RH-2006-Renshaw} attempt to improve estimation efficiency by simplifying the structure of the Renshaw-Haberman model. Specifically, they consider setting $c_x$ in the original Renshaw-Haberman Model to $1/p$, where $p$ represents the number of ages covered by the data set. The resulting model, given by
\begin{equation}\label{H1}
    y_{x,t}:=\log(m_{x,t})=a_x+b_xk_t+\frac{1}{p}\gamma_{t-x}+\varepsilon_{x,t},
\end{equation}
is often referred to as the \textit{H1 model}, and is further discussed by \cite{Comparison-2011-Haberman}. The H1 model may be further reduced by setting $b_x=1/p$. This further simplification would result in the classical age-period-cohort (APC) model \citep{Cohort-1982-Hobcraft}:
\begin{equation}\label{APC}
    y_{x,t}:=\log(m_{x,t})=a_x+\frac{1}{p}k_t+\frac{1}{p}\gamma_{t-x}+\varepsilon_{x,t}.
\end{equation}
Reducing the model structure may result in a faster convergence; however, a reduced model structure may no longer provide an adequate fit. 

\subsubsection{The Hunt-Villegas Method}\label{HV}

\cite{RH-2015-Hunt} argue that the slow convergence of the MLE for the Renshaw-Haberman model is due possibly to an approximate identifiability issue. 

Specifically, \cite{RH-2015-Hunt} show that if $k_t$ in \eqref{RH} follows a perfect straight line, then there exists an approximately invariant parameter transformation. In other words, parameters are not unique even if the four parameter constraints specified in \eqref{LC-constraints} and \eqref{RH-add-constraints} are imposed. 

Empirically, the estimates of $k_t$ typically exhibit a steady downward trend due to mortality improvements, but the trend is not perfectly linearly. As such, this identification problem is `approximate' rather than `exact'. The approximate identification problem means that there exist different sets of parameters that would lead to different allocations between the time effect and cohort effect but approximately the same fit to the historical data. This phenomenon could potentially make the optimization procedure slow and unstable.

To resolve the approximate identifiability issue, \cite{RH-2015-Hunt} suggest imposing an additional constraint:
\begin{equation}\label{RH-Appro-Constr-2015}
    \sum_{s=t_1-x_p}^{t_n-x_1}(s-\bar{s})\gamma_s=0,
\end{equation}
where $\bar{s}$ represents the average year-of-birth over the years-of-birth covered by the data set. This constraint ensures that $\gamma_s$ does not follow a linear trend over the years-of-birth covered by the data set. To see why this is true, we can treat \eqref{RH-Appro-Constr-2015} as a requirement that the sample covariance between $\gamma_s$ (which has a zero mean due to another identifiability constraint) and year-of-birth $s$ is zero. \cite{Mortality-2021-Hunt} mention that the additional constraint has significant demographic significance. Specifically, it is conceivable that $\gamma_s$ is approximately trendless, because systematic changes in mortality over time should have been captured by  $k_t$. 

Imposing the additional constraint can mitigate the approximate identification issue. It also shrinks the parameter space over which the Newton-Raphson's algorithm has to cover, thereby stabilizing and accelerating the optimization.


\section{The Proposed Method}

\subsection{Motivation}

The existing methods for expediting Renshaw-Haberman estimation are both based on the MLE framework, and therefore the requirement of a distributional assumption (which may turn out to be wrong) remains. Also, both methods rely on a reduction in parameter space, so that the improve estimation efficiency at the expense of goodness-of-fit.  

The aforementioned limitations motivate us to tackle the estimation challenge from a different angle. Specifically, we develop a least squares approach for the Renshaw-Haberman model, in which computational efficiency is achieved through some closed-form SVD solutions. The proposed approach has the following merits:

\begin{itemize}
    \item {\it A sharper objective function} 
    
    Compared to MLE, the proposed least squares method is based on a different objective function. We show empirically in Section \ref{sec:NA} that the objective function in our proposed method is sharper, thereby resulting in a faster convergence. Also, the objective function is optimized in part by some SVD closed-form solutions, so that our proposed method is more computational efficient. 
        
    \item {\it Free of any distributional assumption}
    
    The MLE approach requires a distributional assumption. The commonly used Poisson death count assumption is not without criticism. For instance, the over-dispersion problem arising from population heterogeneity would render the Poisson assumption inappropriate. Although this problem may be mitigated by assuming a more flexible death count distribution, such as negative binomial \citep{li2009uncertainty}, the number of parameter would increase and consequently model estimation may be even slower. In contrast, the least squares method we propose requires no distribution assumption, as it obtains parameter estimates by directly minimizing the sum of squared differences between observed and fitted log mortality rates. 

    \item {\it Seamless integration with existing methods for expediting estimation}
    
    The proposed estimation method applies to not only the original Renshaw-Haberman model but also its reduced versions including the H1 model. It can also be implemented with the Hunt-Villegas method to further improve computational efficiency.
    
\end{itemize}

\subsection{Main Optimization: Alternating Minimization}

When a least squares approach is used to estimate the Renshaw-Haberman model, the optimization problem can be expressed as
\begin{equation}\label{RH-MSE}
\min_{\bm{a},\bm{b},\bm{k},\bm{c},\bm{\gamma}}\sum_{x,t}(y_{x,t}-(a_x+b_xk_t+c_x\gamma_{t-x}))^2.
\end{equation}
The parameter constraints specified in \eqref{LC-constraints} and \eqref{RH-add-constraints} are imposed to stipulate a unique solution. 

Unlike the the least squares optimization problem \eqref{Lee-Carter MSE} for the Lee-Carter model,  \eqref{RH-MSE} is a much more challenging non-convex problem that cannot be easily solved. To overcome the estimation challenge, we can consider an \textit{alternating minimization} strategy, in which the shape parameter vector $\bm{a}$, the age-period component $(\bm{b},\bm{k})$ and the age-cohort component $(\bm{c},\bm{\gamma})$ are updated in turns, while other components are held fixed. 

\begin{algorithm}[t!]
\begin{enumerate}
    \item Set initial values of $\bm{\theta}:=(\bm{a},\bm{b},\bm{k},\bm{c},\bm{\gamma})$. 

    \item Update $\bm{a}$ while $\bm{b}$, $\bm{k}$, $\bm{c}$ and $\bm{\gamma}$ are fixed:
    \begin{equation}\label{RH-Estimation-a}
        \min_{\bm{a}}\sum_{x,t}(\underbrace{(y_{x,t}-b_xk_t-c_x\gamma_{t-x})}_{\text{given}}-a_x)^2.
    \end{equation}
    This sub-optimization is accomplished with the following explicit solution: 
    \begin{equation}\label{RH-Solution-a}
        a_x:=\frac{1}{n}\sum_{t=t_1}^{t_n}\left(y_{x,t}-b_xk_t-c_x\gamma_{t-x} \right)=\frac{1}{n}\sum_{t=t_1}^{t_n}\left(y_{x,t}-c_x\gamma_{t-x} \right).
    \end{equation}
    The last step is the above originates from the identification constraint $\sum_{t=t_1}^{t_n}k_t=0$.

    \item Fixing $\bm{a}$, $\bm{c}$ and $\bm{\gamma}$, update $\bm{b}$ and $\bm{k}$:
    \begin{equation}\label{RH-Estimation-bk}
        \min_{\bm{b},\bm{k}}\sum_{x,t}(\underbrace{(y_{x,t}-a_x-c_x\gamma_{t-x})}_{\text{given}}-b_xk_t)^2.
    \end{equation}
    This sub-optimization is accomplished by applying a first-order SVD to the matrix of $y_{x,t} - a_x - c_x \gamma_{t-x}$.

    \item Update $\bm{c}$ and $\bm{\gamma}$ while $\bm{a}$, $\bm{b}$ and $\bm{k}$ are fixed:
    \begin{equation}\label{RH-Estimation-cgamma}
        \min_{\bm{c},\bm{\gamma}}\sum_{x,t}(\underbrace{(y_{x,t}-a_x-b_xk_t)}_{\text{given}}-c_x\gamma_{t-x})^2,
    \end{equation} 
    This sub-optimization is accomplished by an \textit{iterative SVD algorithm}, described in Section 4.2. Then, the estimates of $\bm{a}$ and $\bm{\gamma}$ are adjusted so that the identifiability constraint $\sum_{s=t_1-x_p}^{t_n-x_1}\gamma_{s}=0$ is satisfied:
    \begin{equation}\label{RH-Adjust gamma}
        \bm{\gamma}:=\bm{\gamma}-\Bar{\bm{\gamma}}, \quad \bm{a}:=\bm{a}+\bm{c}\Bar{\bm{\gamma}}, \quad \text{where }\Bar{\bm{\gamma}}=\frac{1}{n+p-1}\sum_{s=t_1-x_p}^{t_n-x_1}\gamma_{s}.
    \end{equation}

    \item Repeat Steps 2-4 until the convergence criterion is satisfied. 
    \end{enumerate}
\caption{The main iteration}
\end{algorithm}

The core iteration is outlined in Algorithm 1. Each cycle of iteration is composed of several steps. Step 2 is simple, as it updates the shape vector $\bm{a}$ through an explicit averaging formula. Step 3 is also straightforward as it just fits a Lee-Carter structure, through a SVD, to the residual after the removal of the shape vector and age-cohort effects. Note that the updated values of $k_t$ from Step 3 sum to zero, because the input residual matrix $[y_{x,t}-a_x-c_x\gamma_{t-x}]_{x,t}$ is row-centered following the implementation of \eqref{RH-Estimation-a} in Step 2.

Step 4, however, represents a much more complex optimization challenge, because it cannot be directly translated into a traditional PCA problem. Therefore, efficiently solving \eqref{RH-Estimation-cgamma} in Step 4 is a crucial milestone of our research question. In the next subsection, we show that Step 4 can be formulated as a PCA problem with missing values, which can be efficiently solved in an iterative manner. 

Finally, Step 5 requires a convergence criterion. In this paper, convergence is achieved when the relative change in the objective function, as defined by \eqref{RH-MSE}, falls below a pre-determined small threshold. Algorithm 1 always converges, since each of Steps 2-4 in the algorithm consistently decreases the objective function, and the objective function ($L^2$ error) is inherently bounded below by zero. 

\subsection{Updating $\bm{c}$ and $\bm{\gamma}$: PCA with Missing Values via an Iterative SVD}

In this subsection, we develop a method to overcome the optimization challenge in Step 4 of Algorithm 1.

First, let us explain in more detail the optimization challenge we are facing. Let $z_{x,t}:=y_{x,t}-a_x-b_xk_t$ be the input for the sub-optimization problem in Step 4. We may arrange the values of $z_{x,t}$ in a $p\times n$ age-period (age-time) matrix as follows:
\begin{equation}\label{Age-Period matrix}
\bm{Z}_{ap} := \begin{bmatrix}
z_{x_1,t_1} & z_{x_1,t_2}    & \cdots & \cdots      & z_{x_1,t_{n-1}} & z_{x_1,t_n}   \\
z_{x_2,t_1} & z_{x_2,t_2}    & \cdots & \cdots      & z_{x_2,t_{n-1}} & z_{x_2,t_n}   \\
\vdots  & \vdots   &         & \iddots     & \vdots  & \vdots       \\
\vdots  & \vdots   & \iddots &             & \vdots  & \vdots        \\
z_{x_{p-1},t_1} & z_{x_{p-1},t_2}    & \cdots & \cdots      & z_{x_{p-1},t_{n-1}} & z_{x_{p-1},t_n} \\
z_{x_p,t_1} & z_{x_p,t_2}    & \cdots & \cdots      & z_{x_p,t_{n-1}} & z_{x_p,t_n}  \\
\end{bmatrix}
\end{equation}
We are unable to update the age-cohort component $(\bm{c},\bm{\gamma})$ by applying a SVD directly to this age-period matrix, because age and cohort are not orthogonal in $\bm{Z}_{ap}$. 

To solve the sub-optimization, we first rearrange the input values in a $p\times (n+p-1)$ age-cohort data matrix:
\begin{equation}\label{Age-Cohort matrix}
   \bm{Z}_{ac}:= \begin{bmatrix}
\times  & \times    & \cdots & \cdots      & \times    & z_{x_1,t_1}& \cdots  & z_{x_1,t_{n-1}} & z_{x_1,t_n} \\
\times  & \times    & \cdots  &\cdots      & z_{x_2,t_1}   & z_{x_2,t_2}& \cdots  & z_{x_2,t_n}& \times \\
\vdots  & \vdots   &       & \iddots      &           &     & \iddots & \vdots & \vdots \\
\vdots  & \vdots    & \iddots &             &        & \iddots &         & \vdots & \vdots \\
\times  & z_{x_{p-1},t_1} & \cdots & z_{x_{p-1},t_{n-1}} & z_{x_{p-1},t_n} & \cdots & \cdots  & \times & \times \\
z_{x_p,t_1} & z_{x_p,t_2}   & \cdots & z_{x_p,t_n}     & \times    & \cdots & \cdots  & \times & \times \\
\end{bmatrix}.
\end{equation}
In $\bm{Z}_{ac}$, each $\times$ represents a missing value, which arise because the oldest and youngest cohorts are not completely observed. For instance, for the youngest cohort of individuals who are born in year $t_n - x_1$, only one observed value $(z_{x_1, t_n})$ is available. In the spirit of this rearrangement, the sub-optimization problem can be expressed as
\begin{equation}\label{RH-Estimation-cgamma-ACscale}
    \min_{\bm{c},\bm{\gamma}}\sum_{x,s\in \mathcal{O}}(z_{x,s}-c_x\gamma_{s})^2,
\end{equation} 
where $s:=t-x$ represents year-of-birth and $\mathcal{O}$ is the set of indices of the observed values. 

If $\bm{Z}_{ac}$ contains no missing value, then we can solve \eqref{RH-Estimation-cgamma-ACscale} readily by applying a SVD to $\bm{Z}_{ac}$. However, given the presence of missing values, the sub-optimization boils down to a first-order PCA with missing values. 

Handling a PCA with missing values is a complex problem in statistics and machine learning \citep{PCA-2010-Ilin}. In the modern statistics and machine learning literature, there exist advanced techniques for handling missing data in PCA, such as matrix completion with a nuclear norm regularization \citep{MatrixComp-2010-Mazumder}. However, such advanced techniques are designed for extremely large-scale and sparse matrices. Additionally, their primary goal is to predict the missing values (matrix completion) rather than finding the optimal least squares solution.

In this study, we utilize a method called the \textit{iterative SVD algorithm}. This algorithm begins with an imputation of the missing values, typically with row-wise means of the observed values in the input matrix. This creates an approximate complete matrix, to which a PCA can be applied to obtain singular vectors (parameter estimates). Then, a PCA reconstruction is employed to generate an improved imputation of the missing values. The process is repeated until convergence is achieved. The implementation of the iterative SVD algorithm in the context of our research is presented in Algorithm 2.

\begin{algorithm}[t]
\begin{enumerate}
    \item Obtain the initial approximate complete matrix $\bm{Z}^*_{ac}$ by imputing the missing values in $\bm{Z}_{ac}$ with row-wise means of the observed values in $\bm{Z}_{ac}$.
 
    \item Apply a SVD to the approximate complete matrix $\bm{Z}^*_{ac}$. Incorporating the identifiability constraint $\sum_{x=x_1}^{x_p}c_x=1$, the updated estimates of $\bm{c}$ and $\bm{\gamma}$ are given by the following expressions
    \begin{equation}\label{Iterative SVD-M}
        \bm{c}:=\frac{\bm{u}_c}{\1^T\bm{u}_c}, \quad \bm{\gamma}:= (\1^T\bm{u}_c) \cdot \bm{Z}_{ac}^{*T} \bm{u}_c,
    \end{equation}
where $\bm{u}_c$ is the first left-singular vector of the approximate complete matrix $\bm{Z}^*_{ac}$. Note that the constraint $\sum_{t-x}\gamma_{t-x}=0$ is incorporated in Algorithm 1 through \eqref{RH-Adjust gamma}.
    
    \item Update the missing values in $\bm{Z}_{ac}$ by a PCA reconstructions with the estimates of $\bm{c}$ and $\bm{\gamma}$ obtained from Step 2. In particular, the missing values in $\bm{Z}_{ac}$ are imputed as $\bm{c}\bm{\gamma}^T$, while the observed values in $\bm{Z}_{ac}$ remain unchanged.

    \item Repeat Steps 2 and 3 until the relative change in the objective function specified by \eqref{RH-Estimation-cgamma-ACscale} is smaller than a certain pre-determined tolerance level.
\end{enumerate}
\caption{Iterative SVD Algorithm}
\end{algorithm}

While the iterative SVD algorithm appears to be a suitable method for solving PCA with missing values, it is not immediately clear why this algorithm addresses our specific $L^2$ minimization problem with missing values. To elucidate this, in the Appendix, we prove that the iterative SVD algorithm minimizes the target loss function specified in \eqref{RH-Estimation-cgamma-ACscale}, and that the iterative SVD algorithm always converges. 


\section{Integrating the Proposed Method with the Existing Methods}

We may implement our proposed method with one or both of the the existing methods (H1 and Hunt-Villegas) to further boost estimation speed. 

\subsection{Implementing with the H1 Model}

Our proposed method can be applied to the variants of the Renshaw-Haberman model, including the H1 model discussed in Section 3. For the H1 model, least squares estimation can be formulated as the following the optimization problem:
\begin{equation}\label{H1-MSE}
\min_{\bm{a},\bm{b},\bm{k},\bm{\gamma}}\sum_{x,t}\left(y_{x,t}-(a_x+b_xk_t+\frac{1}{p}\gamma_{t-x})\right)^2,
\end{equation}
and the following three identification constraints can be used to stipulate parameter uniqueness:
\begin{equation}\label{H1-constraints}
    \sum_{x=x_1}^{x_p}b_x=1, \quad \sum_{t=t_1}^{t_n}k_t=0,\quad
    \sum_{t-x=t_1-x_p}^{t_n-x_1}\gamma_{t-x}=0.
\end{equation}

The main algorithm of our proposed method for the H1 model is identical to Algorithm 1 for the Renshaw-Haberman model, except that $c_x$ is always set to $=1/p$. Interestingly, as explained below, further computational simplifications can be achieved when our proposed method is applied to the H1 model.

For the H1 model, we can update $\bm{\gamma}$ using explicit formulas, thereby eliminating the need for iterative algorithms. To explain, we first express the sub-optimization problem for updating $\bm{\gamma}$ (Step 4 in Algorithm 1) in the H1 model as follows:
\begin{equation}\label{H1-Estimation-gamma}
        \min_{\bm{\gamma}}\sum_{x,t}\left(\underbrace{(y_{x,t}-a_x-b_xk_t)}_{\text{given}}-\frac{1}{p}\gamma_{t-x}\right)^2.
\end{equation} 
The above can be rewritten in an age-cohort dimension as 
\begin{equation}\label{H1-Estimation-gamma-missing}
        \min_{\bm{\gamma}}\sum_{x,s\in \mathcal{O}}\left(z_{x,s}-\frac{1}{p}\gamma_{s}\right)^2,
\end{equation} 
where $z_{x,s}:=y_{x,s}-a_x-b_xk_s$ denotes the residual from Step 3 in Algorithm 1, $s:=t-x$ represents year-of-birth, and $\mathcal{O}$ is the set of the indices for the observed values in $\bm{Z}_{ac}$ (the matrix of $z_{x,s}$ in age-cohort dimension). 

Noticing that \eqref{H1-Estimation-gamma-missing} is separable, we can rewrite it as:
\begin{equation}\label{H1-Estimation-gamma-ACscale-Expanded}
    \min_{\bm{\gamma}}\sum_s  \sum_{x \in \mathcal{O}_s}\left(z_{x,s}-\frac{1}{p}\gamma_{s}\right)^2,
\end{equation}
where $\mathcal{O}_s$ denotes the set of the indices for the observed values in column $s$ of $\bm{Z}_{ac}$. Note that this convenient separability does not hold for the general Renshaw-Haberman model with $c_x\neq 1/p$, since each summand $(z_{x,s}-c_x\gamma_s)^2$ depends on both age $x$ and year-of-birth $s$. 

The separability enables us to solve the target optimization problem \eqref{H1-MSE} by solving the following for each $s$:
\begin{equation}\label{H1-Estimation-gamma-ACscale-Sub}
    \min_{\gamma_s}\sum_{x\in \mathcal{O}_s}\left(z_{x,s}-\frac{1}{p}\gamma_{s}\right)^2.
\end{equation}
For a given $s$, \eqref{H1-Estimation-gamma-ACscale-Sub} is a simple linear regression with no intercept and a slope of 
\begin{equation}\label{H1-gamma-solution}
    \hat{\gamma}_s= \frac{p}{n_s}\sum_{x \in \mathcal{O}_s}z_{x,s},
\end{equation}
where $n_s:=|\mathcal{O}_s|$ is the cardinality of $\mathcal{O}_s$. Applying \eqref{H1-gamma-solution} for every year-of-birth $s$ covered by the data set yields an update of $\bm{\gamma}$.

\subsection{Implementing with the Hunt-Villegas Method}

This subsection explains how our proposed method can be utilized with the H1 model and the Hunt-Villegas method. 

Recall that the Hunt-Villegas method originates from an approximate identifiability problem of the Renshaw-Haberman model and its variants. For the H1 model, \cite{RH-2015-Hunt} show that if $k_t$ follows a perfect straight line, i.g., $k_t = K(t-\bar{t})$, where $K$ is a constant that is less than zero and $\bar{t}=(t_n + t_1)/2$ represents the mid-point of the calibration window, then there exists the following invariant transformation that is equivalent to $\{a_x,b_x,k_t,\gamma_{s}\}$:
\begin{equation}\label{H1-transformation}
    \left\{a_x+\frac{g}{p}(x-\bar{x}),\frac{K}{K-g}b_x-\frac{g}{p(K-g)},\frac{K-g}{K}k_t,\gamma_s +g(s-\bar{s}) \right\},
\end{equation}
where $\bar{x} = (x_p + x_1)/2$ represents the mid-point of the age range under consideration and $g$ is a real constant. In practice, the trend in $k_t$ close to but not perfectly linear, so that an {\it approximate} identifiability problem exists. This approximate identifiability problem may adversely affect convergence of the estimation algorithm. \cite{RH-2015-Hunt} propose to mitigate approximate identifiability problem by imposing the extra constraint specified in \eqref{RH-Appro-Constr-2015}.

\cite{RH-2015-Hunt} proposed a modified Newton-Raphson method to impose \eqref{RH-Appro-Constr-2015} in Poisson ML estimation of model parameters. Specifically, in each iteration of the Newton-Raphson algorithm, they determine the values of $K$ and $g$ in the invariant transformation such that \eqref{RH-Appro-Constr-2015} is satisfied; then, the invariant transformation is applied to adjust the parameter estimates. 

However, it turns out that the modified Newton-Raphson method is not applicable to our alternating minimization scheme. To explain, let us suppose that in one iteration we have updated the value of $\bm{\gamma}$ using the closed-form solution provided in \eqref{H1-gamma-solution}. This update is guaranteed to decrease the value of the overall objective function specified in \eqref{H1-MSE}. However, if we adjust the estimates of $\bm{a}$, $\bm{b}$, $\bm{k}$, and $\bm{\gamma}$ using the approximate invariant transformation specified in \eqref{H1-transformation} to make \eqref{RH-Appro-Constr-2015} hold, then the resulting estimates may lead to a higher (less optimal) value of \eqref{H1-MSE}, since the transformation is only approximately (rather than exactly) invariant. If the value of the objective function increases in some iterations, the alternating minimization algorithm may diverge. 

We propose to incorporate the additional constraint specified in \eqref{H1-transformation} by using a Lagrange multiplier in the update of $\bm{\gamma}$ in model H1. Incorporating \eqref{H1-transformation}, we aim to solve the following constrained optimization problem in the update of $\bm{\gamma}$: 
\begin{equation}\label{RH-Estimation-cgamma-ACscale-Constr}
    \min_{\bm{c},\bm{\gamma}}\sum_{x,s\in \mathcal{O}}(z_{x,s}-c_x\gamma_{s})^2,\quad \text{s.t. } \sum_{s= t_1-x_p}^{t_n-x_1} \gamma_s(s-\bar{s})=0.
\end{equation}
Then, the Lagrangian can be written as:
\begin{equation}\label{H1-Estimation-gamma-ACscale-Lagrangian}
    \mathcal{L}(\bm{\gamma},\lambda)=\sum_{x,s\in \mathcal{O}}\left(z_{x,s}-\frac{1}{p}\gamma_{s}\right)^2+2\lambda\sum_{s= t_1-x_p}^{t_n-x_1}\gamma_s(s-\bar{s}),
\end{equation}
where $\lambda$ represents the Lagrangian multiplier.\footnote{We multiply $\lambda$ by two for computational convenience. The use of $2\lambda$ instead of $\lambda$ makes no difference in the final solution.}

Unlike the unconstrained case in which objective function is separable, the minimization of \eqref{H1-Estimation-gamma-ACscale-Lagrangian} is non-separable because the Lagrange multiplier $\lambda$ applies to all years-of-birth $s = t_1-x_p, \ldots, t_n-x_1$. To obtain the solution to \eqref{RH-Estimation-cgamma-ACscale-Constr}, we derive the first-order partial derivatives of $\mathcal{L}(\bm{\gamma},\lambda)$ with respect to $\bm{\gamma}$ and $\lambda$, and set them to zero:
\begin{equation}\label{Lagrangian-gradient-gamma}
    \frac{\partial \mathcal{L}}{\partial \gamma_s}=-\frac{2}{p}\cdot \sum_{x \in \mathcal{O}_s}\left(z_{x,s}-\frac{1}{p}\gamma_s\right)+2\lambda (s-\bar{s})=0, \quad s= t_1-x_p, \ldots, t_n-x_1. 
\end{equation}
\begin{equation}\label{Lagrangian-gradient-lambda}
    \frac{\partial \mathcal{L}}{\partial \lambda}=\sum_{s= t_1-x_p}^{t_n-x_1} \gamma_s(s-\bar{s})=0. 
\end{equation}
From \eqref{Lagrangian-gradient-gamma}, we obtain the following expression of $\gamma_s$ in terms of $\lambda$:
\begin{equation}\label{Lagrangian-gamma-as-lambda}
    \gamma_s=\frac{p}{n_s}\cdot \left[ \left(\sum_{x \in \mathcal{O}_s}z_{x,s}\right)-p\lambda(s-\bar{s}) \right],
\end{equation}
for $s = t_1-x_p, \ldots, t_n-x_1$. Plugging \eqref{Lagrangian-gamma-as-lambda} into \eqref{Lagrangian-gradient-lambda}, we get the optimal solution for $\lambda$:
\begin{equation}\label{Lagrangian-solution-lambda}
    \hat{\lambda}=\frac{1}{p}\cdot \left[\sum_{s= t_1-x_p}^{t_n-x_1} \frac{(s-\bar{s})^2}{n_s} \right]^{-1} \cdot \sum_{s= t_1-x_p}^{t_n-x_1} \left[\frac{s-\bar{s}}{n_s}\cdot\sum_{x\in \mathcal{O}_s}z_{x,s} \right]. 
\end{equation}
Plugging \eqref{Lagrangian-solution-lambda} back into \eqref{Lagrangian-gamma-as-lambda} gives the solution to $\gamma_s$ for $s = t_1-x_p, \ldots, t_n-x_1$.


\section{Numerical Illustrations}\label{sec:NA}

In this section, we present various experiments to illustrate our proposed least square method for estimating the Renshaw-Haberman model. The data used are obtained from the \cite{HMD-2023-HMD}. They cover a calibration window of 1950-2019 and an age range of 60-89. All of the experiments are performed using a desktop with an Intel Core i9-10900 CPU at 2.80 GHZ, 16 GB of RAM, and Windows 11 Education (64 bits).

All estimation methods under consideration involve an iterative procedure. While it is usual to base the convergence criterion of an iterative procedure on the \textit{absolute} change in the objective function in each iteration, we consider the \textit{relative} change instead, because the objective functions of Poisson ML estimation and the proposed least squares estimation have rather different magnitudes. Basing the convergence criterion on relative changes allows us to compare the two streams of estimation methods more fairly.

We use $\delta$ to represent the tolerance level used in main estimation algorithms. The choice of $\delta$ is admittedly subjective. The \textbf{StMoMo} package, by default, uses a tolerance level of $10^{-4}$ and  the absolute change in the log-likelihood function as the convergence criterion when fitting the Renshaw-Haberman model. Considering the size of the datasets we are using, the values of the maximized log-likelihood functions (when models are fitted using Poisson MLE) have a magnitude of $10^4$. Since we are using basing our convergence criterion on relative changes, the baseline value of $\delta$ is set to $10^{-8}$ to match the standard used in the \textbf{StMoMo} package. 

\subsection{Comparing Least Squares with Poisson ML}

We first compare the following three methods for fitting the Renshaw-Haberman model:
\begin{itemize}
    \item \textit{RH-MLE}: The Renshaw-Haberman model estimated with Poisson MLE;
    \item \textit{RH-MLE-HV}: The Renshaw-Haberman model estimated with Poisson MLE and the Hunt-Villegas method;
    \item \textit{RH-LS}: The Renshaw-Haberman model estimated with our proposed least squares method.
\end{itemize}

The baseline results are obtained using the data from the male populations of England and Wales (E\&W) and the US. These data sets are considered in prominent works on stochastic mortality modelling \citep[e.g.,][]{Comparison-2009-Cairns, Comparison-2011-Cairns}. 

The results are summarized in Tables 1, from which we observe that RH-LS consumes significantly less computation time compared to RH-MLE. Reductions in computational time are over 90\% in general. Figure 1 shows that for a given data set, the parameter estimates from RH-LS and RH-MLE are highly similar. It is not surprising that the parameter estimates from the two estimations methods are not identical, because they are based on different objective functions. As expected, RH-LS (which minimizes the $L^2$ error) yields a lower (less preferred) log-likelihood but a smaller $L^2$ error compared to RH-MLE (see Table 1). 

\begin{figure}[t!]
    \centering
    \includegraphics[width=15cm]{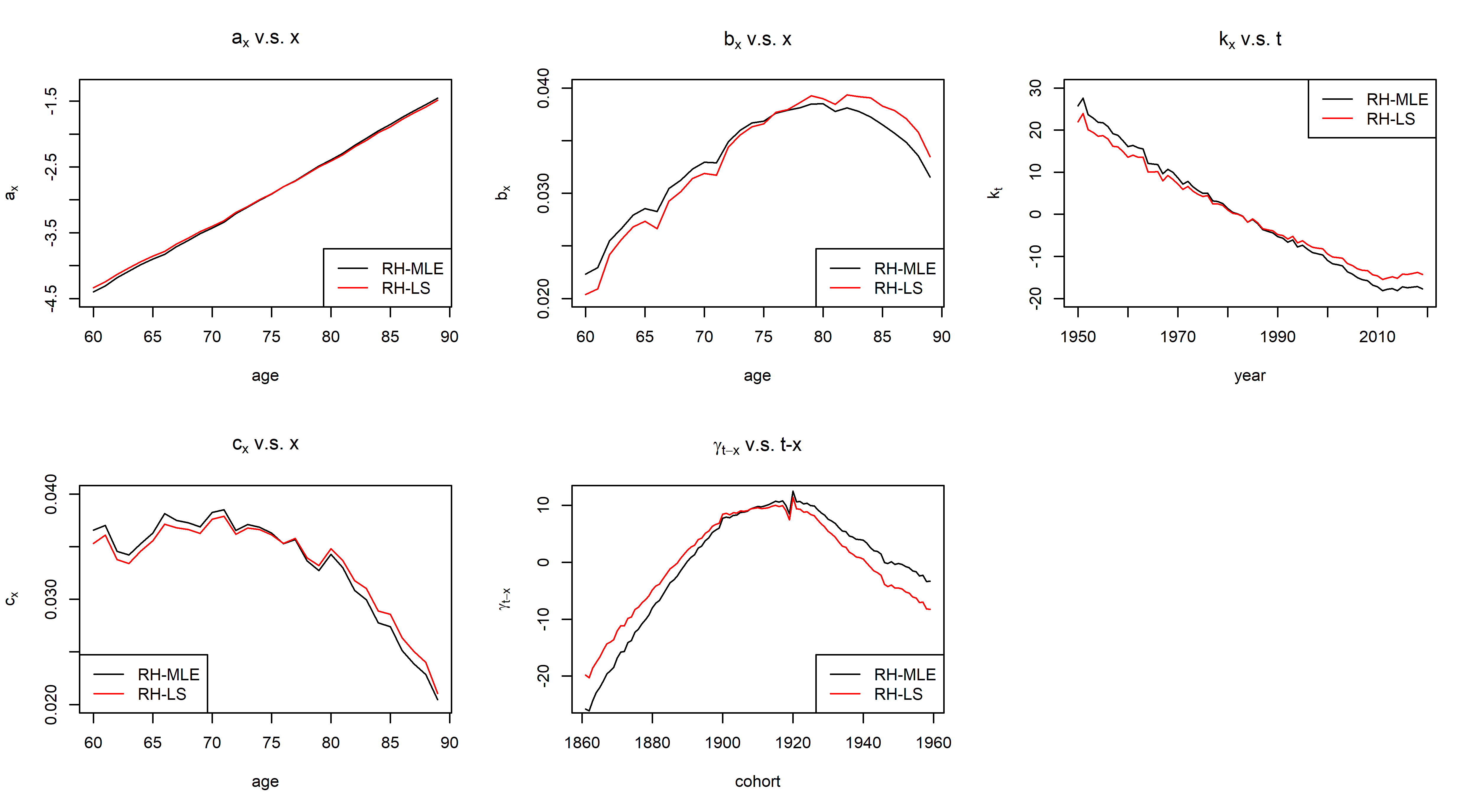}
    \caption{Parameter Estimates derived from the E\&W male dataset, RH-MLE and RH-LS.}
    \includegraphics[width=15cm]{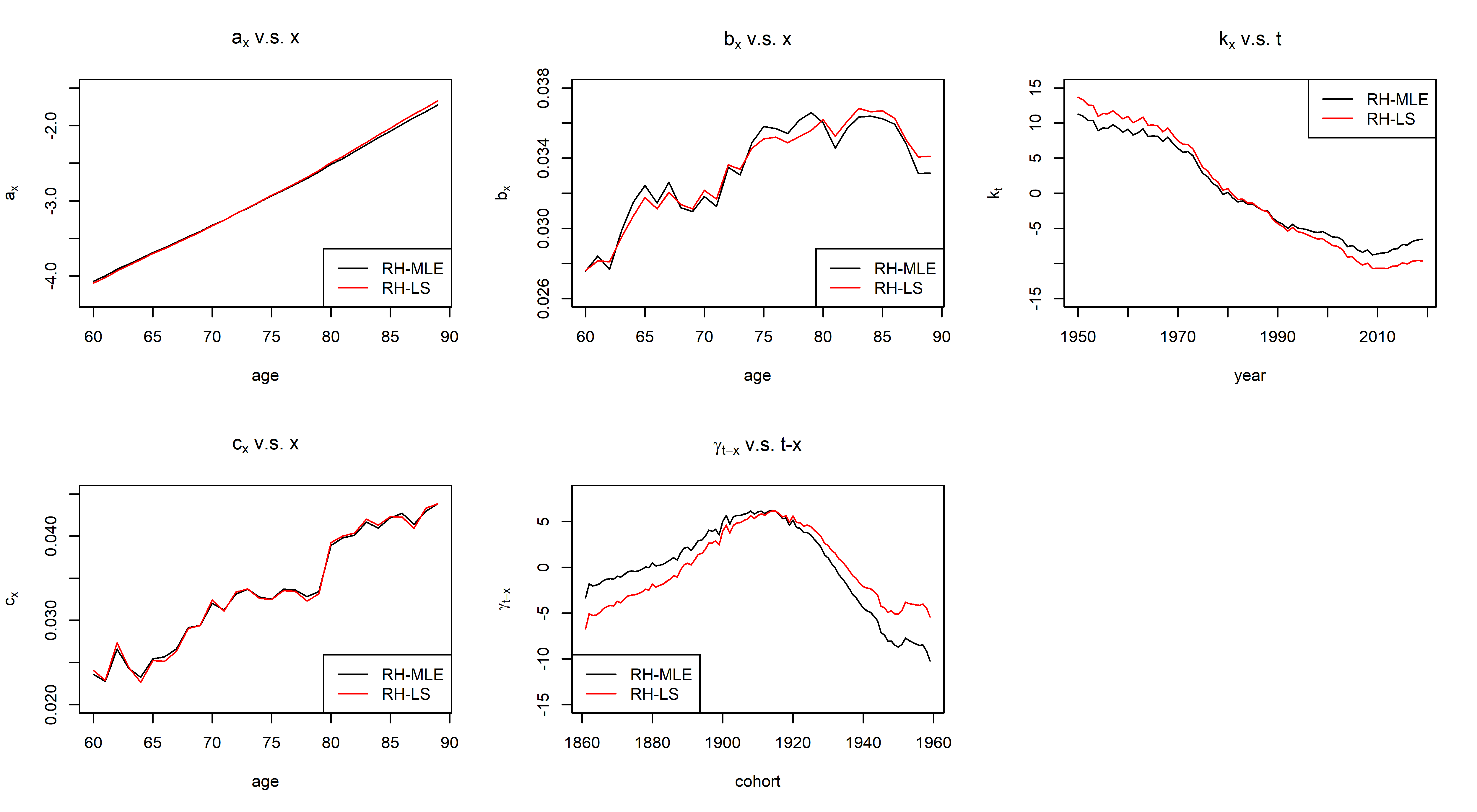}
    \caption{Parameter Estimates derived from the US male dataset, RH-MLE and RH-LS.}
\end{figure}

\begin{table}[h]
\centering
\caption{$L^2$ errors, log-likelihood values, and computation times for RH-MLE and RH-LS, based on E\&W male and US male datasets.}
\begin{tabular}{c c c c c}
\hline \hline
Data &   & \textit{RH-MLE} & \textit{RH-LS} & \textit{RH-MLE-HV}\\ 
\hline

\multirow{3}{*}{E\&W male}   
& $L^2$ error & $0.578$ & $\bm{0.565}$ & $0.581$    \\  
& Log-likelihood & $\bm{-12843}$ & $-12890$ & $-12853$  \\  
& Computing time (seconds) & $336.68$ & $38.72$ & $\bm{20.39}$ \\  
\hline
\multirow{3}{*}{US male}   
& $L^2$ error & $0.472$  & $\bm{0.465}$ & $0.473$  \\  
& Log-likelihood & $\bm{-17736}$  & $-17828$ & $-17744$\\ 
& Computing time (seconds) & $228.72$  & $\bm{10.44}$ & $25.95$ \\ 
\hline \hline

\end{tabular}

\end{table}

\begin{table}[h]
\centering
\caption{Computation times for RH-MLE and RH-LS, based on eight alternative datasets.}
\begin{tabular}{c c c c }
\hline \hline

\multicolumn{4}{c}{Computation time (seconds)}\\ 
\hline

Data & \textit{RH-MLE} & \textit{RH-LS} & \textit{RH-MLE-HV}\\ 
\hline

E\&W female & $132.21$ & $36.28$ & $\bm{10.26}$    \\  
US female & $30.23$ & $\bm{26.18}$ & $42.41$  \\  
Australia male & $53.10$ & $\bm{5.76}$ & $18.28$ \\  
Australia female & $59.12$ & $\bm{10.81}$ & $12.76$ \\  
Canada male & $118.51$ & $\bm{16.17}$ & $29.98$ \\  
Canada female & $59.02$ & $\bm{11.71}$ & $43.31$ \\  
The Netherlands male & $101.22$ & $\bm{13.58}$ & $15.34$ \\  
The Netherlands female & $39.25$ & $24.52$ & $\bm{5.62}$ \\  
\hline \hline
\end{tabular}

\end{table}

To demonstrate the consistency in computation time reduction, we compare RH-LS with RH-MLE using eight alternative data sets: E\&W female, US female, Australia male and female, Canada male and female, and the Netherlands male and female. Reported in Table 2, the results show that RH-LS takes a significantly shorter computation time compared to RH-MLE for all datasets under consideration. While it is more computationally efficient, our proposed method preserves goodness-of-fit in the sense that it results in smaller $L^2$ errors and similar log-likelihood values compared to the Poisson ML approach.

In Tables 1 and 2, we also note that RH-MLE-HV takes shorter computational times relative to RH-MLE, and is comparable to RH-LS in terms of computational efficiency. However, it is important to note that RH-MLE-HV results in less desirable $L^2$ errors and log-likelihoods compared to both RH-MLE and RH-LS, because RH-MLE-HV entails an additional constraint which makes the model more restrictive. That said, RH-MLE-HV improves computational efficiency at the expense of goodness-of-fit. 

\subsection{Sharpness of Objective Functions}

One may wonder why the proposed least squares method is more computationally efficient than the Poisson maximum likelihood approach, while producing a comparable goodness-of-fit. In this sub-section, we attempt to account for the superiority of our proposed approach by considering the sharpness of the objective functions used in each of the candidate estimation methods.

In a study of maximum likelihood estimation of various stochastic mortality models, \cite{Comparison-2009-Cairns} mentioned that ``the likelihood function will be close to flat in certain dimensions.'' As a result of such flatness, over the iterative estimation process, parameter estimates tend stray around the area of parameter space over which the resulting log-likelihood values are similar, thereby resulting in a slow convergence. 

Should there be flatness in certain dimensions of the objective function, parameter estimates tend to be sensitive to the tolerance level $\delta$ used in the iterative estimation process. To compare the sharpness of the objective functions for RH-MLE and RH-LS, we estimate the Renshaw-Haberman model with Poisson MLE and the proposed least squares methods to the US male dataset, for different tolerance levels: $10^{-6}$, $10^{-7}$ and $10^{-8}$. 

Figures 3 reveals that parameter estimates obtained from Poisson MLE are quite sensitive to the tolerance level. The reduction in tolerance level does not materially improve the log-likelihood value, but comes with a substantially longer computation time. In contrast, Figure 4 shows that the parameter estimates obtained from the proposed least squares method are more robust with respect to the tolerance level. Additionally, compared to Poisson MLE, the increase in computational time as the tolerance level reduces is moderate. These outcomes suggest that the objective function for the proposed method is sharper, offering a reason as to why the proposed method is more computationally efficient.

\begin{table}[h!]
\centering
\caption{$L^2$ errors, log-likelihoods, and computing times for RH-MLE and RH-LS when three different tolerance levels are used, US male.}
\begin{tabular}{c c c c}
\hline \hline

 & Tolerance Level & \textit{RH-MLE} & \textit{RH-LS} \\ 
\hline

\multirow{3}{*}{$L^2$ error}   
& $10^{-6}$& $0.4734$ & $0.4661$   \\
& $10^{-7}$ & $0.4727$ & $0.4654$  \\  
& $10^{-8}$ & $0.4720$ & $0.4652$   \\ 

\hline 
\multirow{3}{*}{Log-likelihood}   
& $10^{-6}$ & $-17749$ & $-17828$  \\ 
& $10^{-7}$ & $-17745$ & $-17827$ \\  
& $10^{-8}$ & $-17737$ & $-17827$  \\ 
\hline 
\multirow{3}{*}{Time (seconds)}   
& $10^{-6}$& $12.28$ & $3.96$   \\ 
& $10^{-7}$ & $20.67$ & $4.92$\\  
& $10^{-8}$ & $283.87$ & $11.28$   \\ 
\hline \hline

\end{tabular}
\end{table}

\begin{figure}[t!]
    \centering
    \includegraphics[width=15cm]{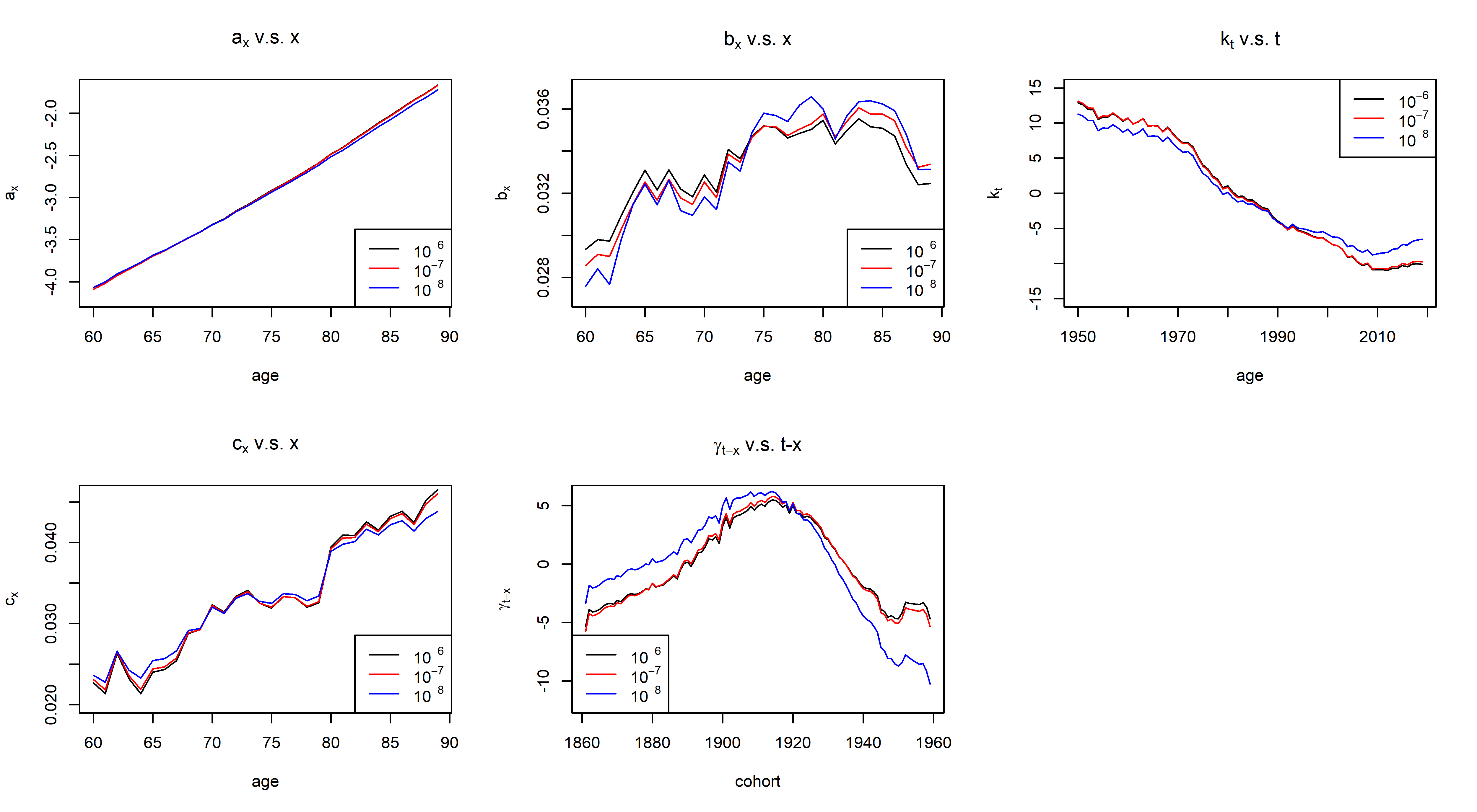}
    \caption{Poisson maximum likelihood estimates of the parameters in the Renshaw-Haberman model for different tolerance levels: $10^{-6}$, $10^{-7}$ and $10^{-8}$; US male dataset.}
    \includegraphics[width=15cm]{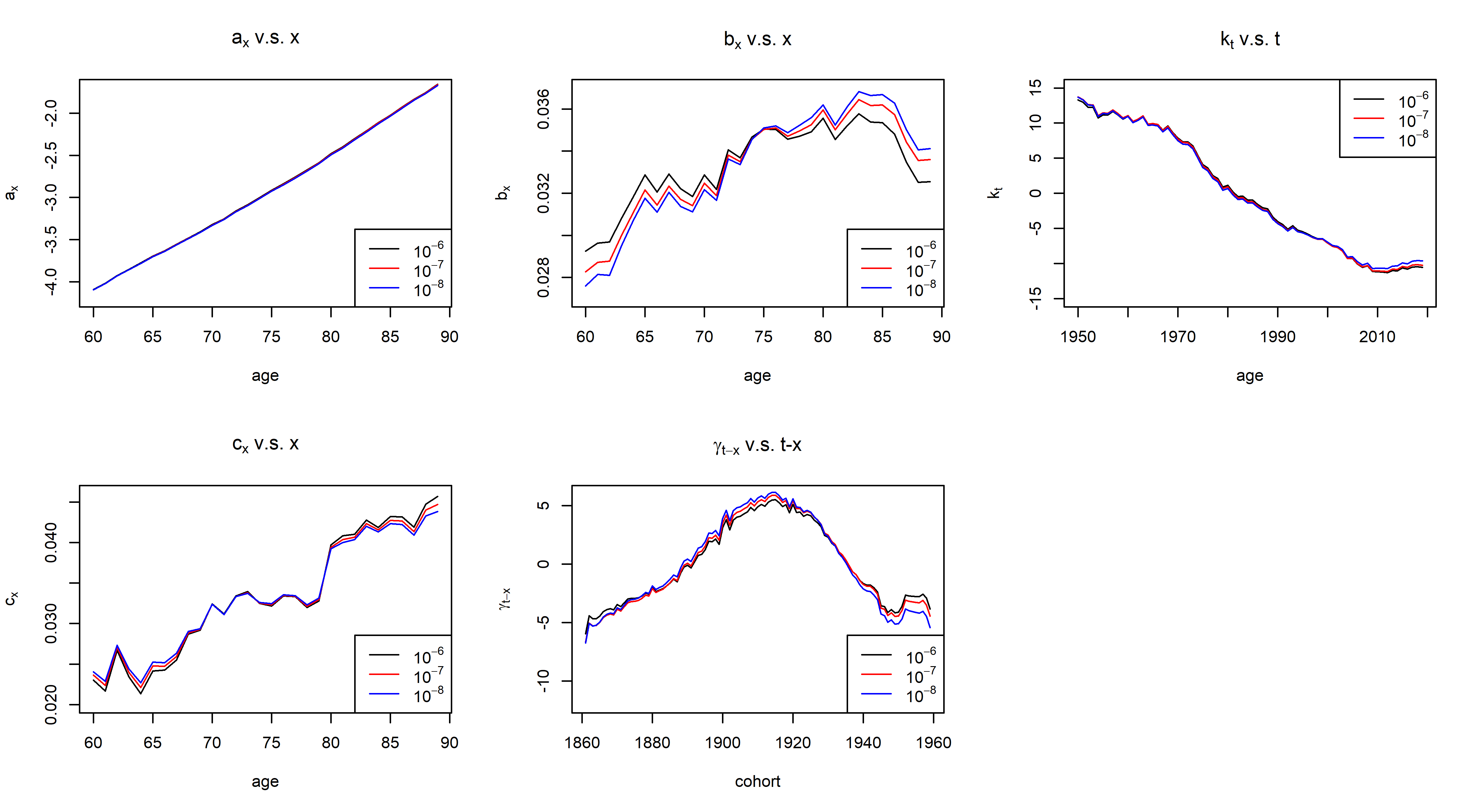}
    \caption{Least squares estimates of the parameters in the Renshaw-Haberman model for different tolerance levels: $10^{-6}$, $10^{-7}$ and $10^{-8}$; US male dataset.}
\end{figure} 

\subsection{Implementing with the H1 Model}

In this sub-section, we implement our proposed least squares estimation method with the H1 model and/or the Hunt-Villegas method to further boost estimation efficiency. The following four settings are considered:
\begin{itemize}

    \item \textit{H1-MLE}: The H1 model estimated with Poisson MLE;
    \item \textit{H1-MLE-HV}: The H1 model estimated with Poisson MLE plus the Hunt-Villegas method;
    \item \textit{H1-LS}: The H1 model estimated with the proposed least squares method;
    \item \textit{H1-LS-HV}: The H1 model estimated with the proposed least squares method plus the Hunt-Villegas method.
    
\end{itemize}

\begin{table}[h]
\centering
\caption{$L^2$ errors, log-likelihoods, and computation times for H1-MLE, H1-MLE-HV, H1-LS, and H1-LS-HV, E\&W male and US male datasets.}
\begin{tabular}{ c c c c c c}
\hline \hline
Data &  & \textit{H1-MLE} & \textit{H1-LS} & \textit{H1-MLE-HV}  & \textit{H1-LS-HV}\\ 
\hline
\multirow{3}{*}{E\&W male}  
& $L^2$ error & $0.682$ & $\bm{0.663}$ & $0.721$  & $0.697$  \\
& Log-likelihood & $\bm{-13149}$ & $-13208$ & $-13247$ & $-13321$  \\ 
& Computation time (seconds) & $62.25$ & $9.79$ & $19.05$ & $\bm{3.16}$  \\ 
\hline
\multirow{3}{*}{US male}  
& $L^2$ error & $0.557$ & $\bm{0.545}$ & $0.557$  & $0.545$  \\
& Log-likelihood & $\bm{-18692}$  & $-18817$ & $-18699$ & $-18816$  \\ 
& Computation time (seconds) & $138.88$  & $2.56$ & $18.30$ & $\bm{2.12}$ \\ 
\hline \hline
\end{tabular}
\end{table}

Table 4 presents the results for the four settings above, derived from the E\&W male and US datasets. Comparing the results for H1-MLE (Table 4) and RH-MLE (Table 3), we notice that using the H1 model (a reduced version of the original Renshaw-Haberman model) helps reduce computation time. Nevertheless, compared to the full Renshaw-Haberman model, the H1 model yields lower log-likelihood values and higher $L^2$ errors, suggesting that it produces a reduced goodness-of-fit. This outcome is expected, as the H1 model is a restricted version of the Renshaw-Haberman model with $x_p$ fewer parameters. 

On the other hand, from Table 4 we observe that the computation times for H1-LS are significantly less than those for H1-MLE, suggesting that the proposed least square estimation methods also offers an improvement in estimation efficiency when a restricted version of the Renshaw-Haberman model is considered. Finally, from Table 4 we notice that H1-LS-HV requires the least computation time among all settings under consideration. For the US male dataset, H1-LS-HV takes just slightly over 2 seconds, which is less than 1\% of the computation time required when we estimate the original Renshaw-Haberman model with Poisson MLE.

For a more comprehensive analysis, we study the four settings with the eight alternative datasets considered in Section 6.1. Tabulated in Table 5, the results indicate the superiority of H1-LS over H1-MLE in terms of computation efficiency for all of the eight datasets under consideration. We also observe from Table 5 that for certain datasets, such as US female, Canada female, and the Netherlands female, fitting the H1 model is very time consuming (even though the H1 model is a restricted version of the original Renshaw-Haberman model), suggesting convergence issues that are possibly caused by the approximate identification problem discussed in Section \ref{HV}. In these cases, using the Hunt-Villegas method could significantly reduce the computation time, and switching from Poisson MLE to the proposed least squares approach could lower the computation time even more. 

\begin{table}[h]
\centering
\caption{Computation times for H1-MLE, H1-MLE-HV, H1-LS, and H1-LS-HV, based on eight alternative datasets.}
\begin{tabular}{c c c c c}
\hline \hline

\multicolumn{5}{c}{Computing time (s)}   \\
\hline

Data & \textit{H1-MLE} & \textit{H1-LS} & \textit{H1-MLE-HV} & \textit{H1-LS-HV} \\ 
\hline
E\&W female & $58.23$ & $1.12$ & $7.58$ & $\bm{0.97}$  \\  
US female & $702.34$ & $89.93$ & $3.58$ & $\bm{1.01}$  \\ 
Australia male & $18.23$ & $7.24$ & $12.09$ & $\bm{2.55}$ \\  
Australia female & $9.12$ & $5.58$ & $6.61$ & $\bm{1.68}$ \\  
Canada male & $18.51$ & $\bm{2.69}$ & $15.23$ & $3.42$ \\  
Canada female & $162.26$ & $78.62$ & $7.28$  & $\bm{2.33}$ \\  
The Netherlands male & $21.22$ & $2.76$ & $9.38$ & $\bm{1.59}$ \\  
The Netherlands female & $425.69$ & $85.51$ & $2.85$ & $\bm{0.36}$ \\  
\hline \hline
\end{tabular}

\end{table}

\section{Concluding Remarks}

In this paper, we introduce a least squares method for estimating the Renshaw-Haberman model. Our proposed approach obtains parameter estimates by minimizing the total $L^2$ error, which measures the sum of squared errors between the observed and fitted log central mortality rates. To overcome the optimization challenge,  we develop an alternating minimization scheme which sequentially updates one group of parameters at a time. We also formulate the update of the age-cohort component as a PCA problem with missing values, so that it can be accomplished effectively using an iterative SVD algorithm. 

Through a number of numerical experiments, we demonstrate that our proposed method significantly outperforms the traditional Poisson MLE in terms of computation time, while producing a better goodness-of-fit in terms of $L^2$ error and a similar goodness-of-fit in terms of log-likelihood. From a theoretical viewpoint, the proposed method does not require any distributional assumption, thereby avoiding the potential problems associated with a distribution mis-specification. 

Our proposed method can be applied to the H1 model, a reduced version of the Renshaw-Haberman model that is designed to improve estimation efficiency. It can also be implemented in tandem with the Hunt-Villegas method, which reduces computation time through an extra parameter constraint. Our numerical experiments indicate that computation time can be reduced further if our proposed method is used with the H1 model and/or the Hunt-Villegas method. 

Future research may extend our proposed estimation method to a wider family of models, including extensions of the Cairns-Blake-Dowd model \citep{Comparison-2009-Cairns}. When \cite{CBD-2006-Cairns} propose the original Cairns-Blake-Dowd model, they estimate it with a least squares approach. However, for the variants of the Cairns-Blake-Dowd model that incorporate cohort effects, least squares estimation is not obvious and future research is needed to investigate the optimization problem.

\section*{Conflict of Interest Statement}

The authors have declared no conflict of interest.


\section*{Appendix: Theoretical Properties of the Iterative SVD Algorithm}

\setcounter{equation}{0}
\renewcommand\theequation{A.\arabic{equation}}

In this appendix, we provide further technical details about the iterative SVD algorithm in Algorithm 2. 

Recall that the objective of the optimization problem is to update the age-cohort parameters $(\bm{c},\bm{\gamma})$ by minimizing the target loss function:
\begin{equation}\label{L_original}
    L(\bm{c},\bm{\gamma})=\sum_{x,s\in \mathcal{O}}(z_{x,s}-c_x\gamma_{s})^2,
\end{equation}
where $s=t-x$ represents year-of-birth and $\mathcal{O}$ is the set of the indexes of the observed values. For notational convenience, we use $\hat{z}_{x,s}(\bm{\theta}):=c_x\gamma_{s}$ to denote the estimator of observation $z_{x,s}$ as a function of the model parameters $\bm{\theta}=(\bm{c},\bm{\gamma})$. We can then regard the optimization as the problem of finding $\bm{\theta}$ that minimizes the $L^2$ loss function of the observed data:
\begin{equation}\label{L_Obs}
    L_{obs}(\bm{\theta})=\sum_{x,s\in \mathcal{O}}(z_{x,s}-\hat{z}_{x,s}(\bm{\theta}))^2.
\end{equation}

The iterative SVD algorithm iteratively performs impute the missing values in and perform SVD on the complete data matrix $\bm{Z}_{ac}$. Let us write the $L^2$ loss function of the complete data and missing data as
\begin{equation}\label{L_comp}
    L_{tot}(\bm{\theta},\Tilde{\bm{z}})=L_{obs}(\bm{\theta})+L_{mis}(\bm{\theta},\Tilde{\bm{z}})
\end{equation}
and
\begin{equation}\label{L_miss}
    L_{mis}(\bm{\theta},\Tilde{\bm{z}})=\sum_{x,s\notin \mathcal{O}}(\Tilde{z}_{x,s}-\hat{z}_{x,s}(\bm{\theta}))^2,
\end{equation}
respectively, where $\Tilde{z}_{x,s}$ is the imputed missing value of $z_{x,s}$. The iterative SVD algorithm minimizes the total $L^2$ error \eqref{L_comp} as a function $L_{tot}(\bm{\theta},\Tilde{\bm{z}})$ with respect to both the model parameter $\bm{\theta}$ and the set of imputed values $\Tilde{\bm{z}}=\{\Tilde{z}_{x,s}|x,s\notin \mathcal{O}\}$. 

\subsection*{Convergence of the Iterative SVD Algorithm}

We first show that the iterative SVD algorithm always converges by showing that the algorithm can be represented as the following alternating minimization procedure:
\begin{enumerate}
    \item For a fixed $\Tilde{\bm{z}}$, let $\bm{\theta}^{\ast}$ be the minimizer of $L_{tot}(\bm{\theta},\Tilde{\bm{z}})$ with respect to $\bm{\theta}$:
    \begin{align}
        \bm{\theta}^{\ast}&=\arg \min_{\bm{\theta}}L_{tot}(\bm{\theta},\Tilde{\bm{z}})\nonumber  \\
        &=\arg \min_{\bm{\theta}}\left[ \sum_{x,s\in \mathcal{O}}(z_{x,s}-\hat{z}_{x,s}(\bm{\theta}))^2+\sum_{x,s\notin \mathcal{O}}(\Tilde{z}_{x,s}-\hat{z}_{x,s}(\bm{\theta}))^2\right]\nonumber \\
        &= \arg \min_{\bm{\theta}}\left[ \sum_{x,s}(z^{\prime}_{x,s}-\hat{z}_{x,s}(\bm{\theta}))^2\right],
    \end{align}
    where $z^{\prime}_{x,s}$ equals $z_{x,s}$ for $x,s\in \mathcal{O}$ and $\Tilde{z}_{x,s}$ for $x,s\notin \mathcal{O}$. Since $\hat{z}_{x,s}(\bm{\theta}):=c_x\gamma_{s}$, the minimizer $\bm{\theta}^{\ast}$ can be found by performing a PCA to the approximate complete matrix $\bm{Z}_{ac}$, as described in Step 2 of Algorithm 2.

    \item For a fixed $\bm{\theta}$, let $\Tilde{\bm{z}}^{\ast}$ be the minimizer of $L_{tot}(\bm{\theta},\Tilde{\bm{z}})$ with respect to $\Tilde{\bm{z}}$:    
    \begin{align}
        \Tilde{\bm{z}}^{\ast}&=\arg \min_{\Tilde{\bm{z}}}L_{tot}(\bm{\theta},\Tilde{\bm{z}})\nonumber \\
        &=\arg \min_{\Tilde{\bm{z}}}\left[L_{mis}(\bm{\theta},\Tilde{\bm{z}})+\text{constant}\right]\nonumber \\
        &= \arg \min_{\Tilde{\bm{z}}}\left[\sum_{x,s\notin \mathcal{O}}(\Tilde{z}_{x,s}-\hat{z}_{x,s}(\bm{\theta}))^2+\text{constant}\right].
    \end{align}
    It is easy to see that the minima is achieved when $\Tilde{z}_{x,s}=\hat{z}_{x,s}(\bm{\theta})$ and so
    $\Tilde{\bm{z}}^{\ast}=\{\Tilde{z}_{x,s}(\bm{\theta})|x,s\notin \mathcal{O}\}$. This solution is exactly the same as imputing the missing values by the PCA reconstruction $\hat{z}_{x,s}(\bm{\theta})$ with parameters $\bm{\theta}$, as described in Step 3 of Algorithm 2.
\end{enumerate}

\subsection*{The Iterative SVD Algorithm Minimizes the Target Loss Function}
We next show that the iterative SVD algorithm minimizes the target loss function \eqref{L_Obs}. More precisely, the minimizer $\bm{\theta}^{\ast}$ obtained by iteratively minimizing the total $L^2$ error $L_{tot}(\bm{\theta},\Tilde{\bm{z}})$ is equivalent to the one obtained by directly minimizing the $L^2$ error $L_{obs}(\bm{\theta})$ of the observed data. 

Following (A.6), we have
\begin{equation}\label{z_imp^ast}
    \Tilde{\bm{z}}^{\ast}=\arg \min_{\Tilde{\bm{z}}}L_{tot}(\bm{\theta},\Tilde{\bm{z}})=\arg \min_{\Tilde{\bm{z}}}\left[\sum_{x,s\notin \mathcal{O}}(\Tilde{z}_{x,s}-\hat{z}_{x,s}(\bm{\theta}))^2\right],
\end{equation}
which immediately implies that
\begin{equation}\label{partial=0}
    \frac{\partial L_{tot}}{\partial \Tilde{\bm{z}}}\Bigg|_{\Tilde{\bm{z}}=\Tilde{\bm{z}}^{\ast}}=0
\end{equation}
and
\begin{equation}\label{C_mis=0}
    L_{mis}(\bm{\theta},\Tilde{\bm{z}}^{\ast})=\sum_{x,s\notin \mathcal{O}}(\hat{z}_{x,s}(\bm{\theta})-\hat{z}_{x,s}(\bm{\theta}))^2=0.
\end{equation}
Therefore, we can obtain
\begin{equation}\label{L_tot=L_mis}
    L_{tot}(\bm{\theta},\Tilde{\bm{z}}^{\ast})=L_{obs}(\bm{\theta})+L_{mis}(\bm{\theta},\Tilde{\bm{z}}^{\ast})=L_{obs}(\bm{\theta}),
\end{equation}
and consequentially, 
\begin{equation}\label{d=lambda}
    \frac{dL_{obs}}{d\bm{\theta}}=\frac{dL_{tot}}{d\bm{\theta}}\Bigg|_{\Tilde{\bm{z}}=\Tilde{\bm{z}}^{\ast}}=\frac{\partial L_{tot}}{\partial\bm{\theta}}\Bigg|_{\Tilde{\bm{z}}=\Tilde{\bm{z}}^{\ast}}+\underbrace{\frac{\partial L_{tot}}{\partial \Tilde{\bm{z}}}\Bigg|_{\Tilde{\bm{z}}=\Tilde{\bm{z}}^{\ast}}}_{=0 \text{ from }\eqref{partial=0}} \cdot \frac{\partial \Tilde{\bm{z}}}{\partial \bm{\theta}}= \frac{\partial L_{tot}}{\partial\bm{\theta}}\Bigg|_{\Tilde{\bm{z}}=\Tilde{\bm{z}}^{\ast}},
\end{equation}
which suggests that the minimizer $\bm{\theta}^{\ast}$ of $L_{tot}(\bm{\theta},\bm{z}_{imp})$ coincides with the minimizer of $L_{obs}(\bm{\theta})$, and thus shows that the iterative SVD algorithm implicitly solves the optimization problem specified in \eqref{RH-Estimation-cgamma-ACscale}.


\end{document}